\documentclass[journal]{IEEEtran}

\ifCLASSINFOpdf
\else
   \usepackage[dvips]{graphicx}
\fi
\usepackage{url}

\usepackage{graphicx}
\usepackage{algorithm}
\usepackage{algorithmic}
\usepackage{amsmath,amssymb,amsfonts}
\usepackage{array}
\usepackage{bm}
\usepackage{cite}
\usepackage[varg]{txfonts}
\usepackage{wrapfig}
\usepackage{color}
\usepackage{lineno}

\newcommand{\maximize}{\mathop{\rm maximize}\limits}

\newcommand*{\reviewerA}[1]{{#1}}%English BlueGreen
\newcommand*{\reviewerC}[1]{{#1}}%Reviewer-2 (3rdrevise) Red
\begin{document}

\title{Randomized Subspace Newton Convex Method Applied to Data-Driven Sensor Selection Problem}

\author{Taku Nonomura, \IEEEmembership{nonmember, IEEE}, Shunsuke Ono, \IEEEmembership{Member, IEEE},\\
        Kumi Nakai, \IEEEmembership{nonmember, IEEE}, Yuji Saito, \IEEEmembership{nonmember, IEEE}
%\thanks{This work was supported in part by JST, and Japan. }
\thanks{T. Nonomura is with Tohoku University, Sendai, 980-8579, Japan (e-mail: nonomura@aero.mech.tohoku.ac.jp), 
S. Ono is with Tokyo Institute of Technology, Yokohama, 80305 Japan (e-mail:ono@c.titech.ac.jp ),
K. Nakai is Tohoku University, Sendai, 980-8579, Japan (e-mail: nakai@aero.mech.tohoku.ac.jp), 
Y. Saito is with Tohoku University, Sendai, 980-8579, Japan (e-mail: saito@aero.mech.tohoku.ac.jp).}
}

\markboth{Journal of \LaTeX\ Class Files}
{Shell \MakeLowercase{\textit{et al.}}: Bare Demo of IEEEtran.cls for IEEE Journals}
\maketitle
%\linenumbers
\begin{abstract}
The randomized subspace Newton convex methods for the sensor selection problem are proposed. The randomized subspace Newton algorithm is straightforwardly applied to the convex formulation, and the customized method in which the part of the update variables are selected to be the present best sensor candidates is also considered.  In the converged solution, almost the same results are obtained by original and randomized-subspace-Newton convex methods. As expected, the randomized-subspace-Newton methods require more computational steps while they reduce the total amount of the computational time because the computational time for one step is significantly reduced by the cubic of the ratio of numbers of randomly updating variables to all the variables. The customized method shows superior performance to the straightforward implementation in terms of the quality of sensors and the computational time. 
\end{abstract}


\IEEEpeerreviewmaketitle

\section{Introduction}

\IEEEPARstart{T}{he} sensor selection problems gather attention for the distributed parameter systems such as fluid dynamics. A sensor selection problem is defined as an NP-hard combinatorial problem. %For instance, when the brute-force search is conducted for 25 sensor selection from 100 sensor candidates,  $\tiny{ \left( \begin{array}{c}100\\ 25 \end{array}\right)}\approx 2.4\times 10^{23}$ patterns should be investigated. It is impossible to conduct such a direct search approach even in this small sensor problems. Therefore, computationally efficient ways selecting the sensors should be considered. 
Thus far, several relaxed computational methods are proposed for sensor selection even in the linear observation problems. 
Joshi and Boyd\cite{joshi2009sensor} defined convex relaxation formulations of the problems and solved by a Newton method. Here, their approach is called the original convex method, hereafter. On the other hand, the greedy methods based on the determinant of the matrix of the pseudo inverse operation were proposed\cite{manohar2018data,saito2019determinantbased,saito2020data,nakai2020effect}, while its quality is slightly worse than that of the sensor selected by the original convex method of Joshi and Boyd in the case of oversampled sensors. Both original convex and greedy methods previously proposed have been implemented and extended to several applications.\cite{clark2018greedy,manohar2018optimal,astrid2008missing,barrault2004empirical,chaturantabut2010nonlinear,carlberg2013gnat,liu2016sensor,yamada2020fast} 

The original convex method might be preferred because it usually works better than the greedy method. In addition, if a more complicated sensor problem is considered, the original convex method can be flexibly applied to such a problem by constructing the objective function and the constraints of the problem. However, the clear drawback of the convex relaxation method is its computational complexity of $\mathcal{O}(n^3)$ when it is applied to a many-degree-of-freedom problem such as data-driven sensor selection of the fluid dynamics, where $n$ is a total number of sensor candidates. If 
a number of the sensor candidates becomes 1,000,000 as is often the case in the fluid dynamics, the use of the original convex method might be almost impossible for the single workstation environment. Therefore, fast methods for the optimization of the objective function are required. 

Recently, the randomized methods\cite{motwani1995randomized,halko2011finding} have been applied to convex/nonconvex problems\cite{shinetsu2005silicone,cevher2014convex,huang2014randomized,ono2019efficient} in signal and image processing, resulting in the significant reduction of computational time. The subspace of the variables are randomly chosen and the variables in the subspace are optimized in each step in those methods. Those methods require more time steps for convergence of the objective function, while they significantly reduce the computational complexity in one step. This strategy is recently extended to the subspace Newton method\cite{gower2019rsn}. \reviewerA{In the present study, we propose to apply the randomized algorithm to the convex sensor selection problems.  This is because the computational complexity of $\mathcal{O}(n^3)$ can be significantly reduced by limiting a number of update variables, and the combination of the randomized algorithm with the convex relaxation formulation can become the practical best sensor selection method in large scale problem in the present state. It should be noted that this is the first trial of the combination.}
In the present paper, we describe the formulation of the randomized subspace Newton (RSN) method\cite{gower2019rsn} applied to the convex sensor selection formulation and illustrate the improvement in computational time while keeping its performance. 

This paper is organized as follows: In Section. \ref{sec:probandalg}, numerical algorithms including a customized method are explained. In Section. \ref{sec:res}, the test results and computational time are discussed for the original and randomized convex methods. Finally, Section \ref{sec:con} concludes the paper. 

\section{Problems and Algorithms}
\label{sec:probandalg}
The sensor selection problem, its convex relaxation, its randomized implementation and the customized one are described in this order. 

\subsection{Sensor Selection Problems}
Here, we define the following sensor selection problem:
\begin{align}
\bm{y}=\bm{H}\bm{U}\bm{z},
\end{align}
where $\bm{y} \in \mathbb{R}^{p} $ is the observation vector, $\bm{H} \in \mathbb{R}^{p \times n}$ is the sensor location matrix, $\bm{U} \in \mathbb{R}^{n \times r}$ is the sensor candidate matrix and $\bm{z} \in \mathbb{R}^{r}$ is the latent variable vector. Here, $p$, $n$, and $r$ are \reviewerC{a number of sensors to be selected, a total number of sensor candidates, and a number of latent variables}. The sensor location matrix $\bm{H}$ has unity at the sensor location and zeros at the other locations for each row. As discussed in the previous studies, the function which evaluates the sensor locations using D-optimality is defined using the Fisher matrix as follows:
\begin{align}
&\maximize_{\boldsymbol{\tilde w}} f_\text{org}(\tilde{\bm{w}}) \notag \\
&f_\text{org}(\tilde{\bm{w}})=\log \det \left(\bm{U}^{\mathsf{T}}\bm{D}_{\tilde{\bm{w}}}\bm{U}\right).
\label{eq:orgobj}
\end{align}
Here, $\bm{H}^{\mathsf{T}}\bm{H}=\bm{D}_{\tilde{\bm{w}}}=\text{diag}(\tilde{\bm{w}})$ is a diagonal matrix, and  $\tilde{w}_i\in\{0,1\}$ and $\bm{1}^{\mathsf{T}}\tilde{\bm{w}}=p$.
The diagonal component corresponding to selected and unselected sensors are unity and zero, respectively. 

\subsection{Original Convex Method of Sensor Selection Problems}
Joshi and Boyd\cite{joshi2009sensor} relaxed this combinatorial problem by introducing the diagonal matrix $\bm{D}_{\bm{w}} \approx \bm{D}_{\tilde{\bm{w}}}$, where $\bm{w}$ is a weight vector, the component of which ranges from 0 to 1 for each sensor, and $\bm{D}_{\bm{w}}=\text{diag}(\bm{w})$. The objective function becomes as follows by convex relaxation:
\begin{align}
&\maximize_{\boldsymbol{w}} \log \det \left(\bm{U}^{\mathsf{T}} \bm{D}_{\bm{w}} \bm{U}\right) \notag
\\ & \text{s.t.}\quad \bm{1}^{\mathsf{T}}\bm{w}=p, \quad 0 \le w_i \le 1.
\end{align}
The constraints of range of the weights are included in the objective function as cost terms with taking logarithms of the original function, and the matrix multiplication is reformulated using summation as follows:
\begin{align}
&\maximize_{\bm{w}} f(\bm{w}) \notag \\
&f(\bm{w})=\log \det \left( \sum_{i=1}^n{w_i \bm{u}_i^{\mathsf{T}} \bm{u}_i} \right) + \kappa \sum_{i=1}^n{\left(\log({w}_i) + \log(1-{w}_i)\right)}\notag
\\ & \text{s.t.}\quad \bm{1}^{\mathsf{T}}\bm{w}=p, \label{eq:obj}
\end{align}
where $\bm{u}_i$ is the $i$th row vector of $\bm{U}$ and $\kappa$ is a parameter that controls the quality of approximation. This problem has been solved with the Newton iteration. Here, 
\begin{align}
%\delta \bm{w}_\text{n} = - (\nabla^2 f)^{-1}\nabla f
%                         + \left( \frac{\bm{1}^{\mathsf{T}}  (\nabla^2 f^{-1}) \nabla f}{\bm{1}^{\mathsf{T}} (\nabla^2 f)^{-1}\bm{1} }  \right)(\nabla^2 f)^{-1}\bm{1}
\delta \bm{w}_\text{n} = - \bm{G}^{-1}\bm{g}
                         + \left( \frac{\bm{1}^{\mathsf{T}} \bm{G} \bm{g} }{\bm{1}^{\mathsf{T}} \bm{G}^{-1}\bm{1} }  \right)\bm{G}^{-1}\bm{1},
\end{align}
where
\begin{align}
\bm{g}&=\nabla f,\\
\bm{G}&=\nabla^2 f,\\
(\nabla f)_i&= \bm{u}_i\bm{W}^{-1}\bm{u}^{T}_i
                 +{\frac{\kappa}{w_i}-\frac{\kappa}{1-w_i}},\\    
(\nabla^2 f)_{ij}&= \bm{u}_i \bm{W}^{-1}  \bm{u}_j^{T} \bm{u}_j \bm{W}^{-1}\bm{u}^{T}_i
                 -\delta_{ij}{\left(\frac{\kappa}{w_i^2}-\frac{\kappa}{(1-w_i)^2}\right)},\\
\bm{W}&=\bm{U}^{\mathsf{T}} \bm{D}_{\bm{w}} \bm{U}.
\end{align}
A backtracking line search is employed to set a step size $\Delta s$ with keeping the constraints on $\bm{w}$ in (\ref{eq:obj}) ,  and $\bm{w}$ is updated by $\bm{w} + \Delta s \delta \bm{w}$. The Newton iteration is stopped when decrement $(-\bm{g} \delta \bm{w})^{1/2}$ becomes small as in the original paper. 
Once the solution is obtained, the sensors with $k$-largest weights are selected in this procedure. The complexity of this algorithm is $\mathcal{O}(n^3)$ and it costs so much for the recent data-driven sensor selection problems.

\subsection{Randomized Subspace Newton Convex Method}
Recently, the RSN method is proposed.\cite{gower2019rsn} The variables are projected to the randomized space with lower dimension and optimized in each step. This method requires more computational steps while it significantly reduces the computational cost of one step, and it reduces the total computational costs in total. 

The random sketching matrix $\bm{S} \in \mathbb{R}^{n \times s}$ is employed where $s$ is the sketch size. $\bm{S}$ can be a Gaussian random matrix, a sparse random matrix, a Bernoulli random matrix or other, while the random permutation matrix is employed and the computational costs are simply reduced in the present study as follows: 
\begin{align}
\bm{S}_k=\bm{S}_{{s}\text{-random}},
\label{eq:S1}
\end{align}
where $\bm{S}_k$ is the random sketching matrix in the $k$th step and $\bm{S}_{{s}\text{-random}}$ is the random permutation matrix. In this case 
the randomly selected component is unity and the others are zeros for each row of $\bm{S}_{{s}\text{-random}}$, and the selection is not overlapped for any other rows of $\bm{S}_{{s}\text{-random}}$.

The randomized subspace Newton iteration with constraint can be rewritten as follows:
\begin{align}
%\delta \bm{w}_\text{n} = - (\nabla^2 f)^{-1}\nabla f
%                         + \left( \frac{\bm{1}^{\mathsf{T}}  (\nabla^2 f^{-1}) \nabla f}{\bm{1}^{\mathsf{T}} (\nabla^2 f)^{-1}\bm{1} }  \right)(\nabla^2 f)^{-1}\bm{1}
\delta \bm{w}_\text{n} = - \bm{S}_k(\bm{S}_k^{\mathsf{T}}\bm{G}\bm{S}_k)^{-1}\bm{S}_k^{\mathsf{T}}\bm{g}
                         + \left( \frac{\bm{1}^{\mathsf{T}} \bm{S}_k^{\mathsf{T}} \bm{G} \bm{S}_k \bm{g} }{\bm{1}^{\mathsf{T}} (\bm{S}_k^{\mathsf{T}}\bm{G}\bm{S}_k )^{-1}  \bm{1} }  \right) (\bm{S}_k^{\mathsf{T}} \bm{G} \bm{S}_k)^{-1} \bm{1}
                         \label{eq:rsn}
\end{align}
This operator just corresponds to the random selection of updating weights and the optimization of selected weights in each step. The convergence criteria is redefined whether the decrement condition is satisfied in $n/s$ consecutive steps, and the degradation of the results by the incorrect convergence judgment due to the randomness is avoided. The speed up of this algorithm is expected to be $(s/n)^3$ for one step, while more of steps (approximately $n/s$ times more steps) will be required for the convergence. This method is called the RSN convex method, hereafter.

\subsection{Customized Randomized Subspace Newton Convex Method}
In the present problem, clearly, sensors with the larger weights in each step has higher possibility to become a part of the optimal set of sensors than those with smaller weights. \reviewerA{Therefore, we choose $\rho s$ elite sensors with $\rho s$-largest weights and randomly $(1-\rho) s$ sensors for the updates of the weights of sensors, where $0 \le \rho \le1 $ is the ratio for elite sensors in the updated sensors. Basically, $\rho=0.5$ is employed in the present study, unless otherwise mentioned.} This procedure is heuristic, but accelerates the convergence while sorting is required in each step.  Although Eq.\ref{eq:rsn} does not change, the sketch matrix becomes
\begin{align}
\bm{S}_k=[\bm{S}_{\rho s\text{-largest}}^{\mathsf{T}} \quad \bm{S}_{(1 - \rho)s\text{-random}}^{\mathsf{T}}]^{\mathsf{T}},
\label{eq:S2}
\end{align}
where the component corresponding to the sensor location of the $i$th largest weight is unity and the others are zeros in the $i$th row vector of the $\bm{S}_{\rho s\text{-largest}}\in \mathbb{R}^{\rho s\times n}$, and $\bm{S}_{(1-\rho)s\text{-random}} \in \mathbb{R}^{(1-\rho) s\times n}$ is a random permutation matrix. 
The difference between the standard and customized randomized methods is only in the choice of $\bm{S}_k$, and the form of the randomized subspace Newton method in (\ref{eq:rsn}), and convergence criteria do not change. This algorithm requires the computational cost of the `sort' command for the choice the $\rho s$-largest weights, but it is expected to be converged with less steps than standard RSN convex method. This method is called the customized RSN (CRSN) convex method the present paper. Algorithm \ref{alg:rsn} shows the procedure of the RSN and CRSN methods. 

\begin{figure}[!tbp]
\centerline{\includegraphics[width=9cm]{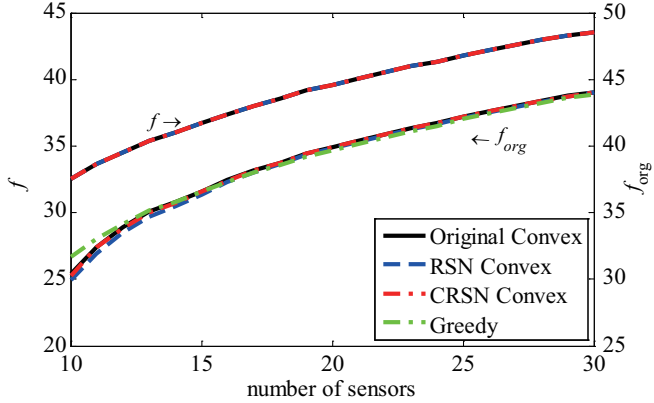}}
\caption{Results of $f$ \reviewerC{o}btained by the convex methods and of $f_{\rm{org}}$ by convex and greedy methods.}
\label{fig:f} 
\end{figure}

\begin{algorithm}[hh]
	\caption{{(C)RSN Convex Sensor Selection Method}} \label{alg:rsn}
	\begin{algorithmic}[1]
		\STATE \textbf{Input:} $\bm{U}$ 
		\STATE \textbf{Output:} $\tilde{\bm{w}}$
		\STATE Set $ \bm{w} = \frac{p}{n}\bm{1} $, $k=1$%
		\WHILE{first step or $(-\bm{g} \delta \bm{w})^{1/2} > \epsilon $  }
		\STATE Set $k \leftarrow k+1$
		\STATE Set fresh random sketch $\bm{S}_k$ as in (\ref{eq:S1}) or (\ref{eq:S2})
		\STATE Calculate $\bm{S}_k \bm{g}$. 
		\STATE Calculate $\bm{w}$ in (\ref{eq:rsn})
		\STATE Obtain $\Delta s$ by backtracking line search
		\STATE Set $\bm{w}=\bm{w}+\Delta s\bm{S}_k\delta\bm{w} $
		\ENDWHILE 
		\STATE Obtain $\tilde{\bm{w}}$ by setting $\tilde{w}_i=1$ for $p$-largest indices of $\bm{w}$ whereas $\tilde{w}_i=0$ for other indices.
	\end{algorithmic}
\end{algorithm}

\section{Results}
\label{sec:res}
The original convex method, the RSN convex method, and the CRSN convex method are compared together with the greedy method as a reference.  
\subsection{Randomized Sensor Selection Problem}
In this section, a randomized sensor selection problem is considered. Numbers of sensor candidates and latent variables, $n$ and $r$, are set to be 10,000 and 10, respectively. This large number of sensor candidates was not considered in the original study by Joshi and Boyd, while it is very time consuming for recent data-driven sensor selection problems as reported by Manohar et al.\cite{manohar2018data} and Saito et al.\cite{saito2019determinantbased}
The number of sensors $p$ are set to be 10, 11, $\dots$ 30. The component of the sensor candidate matrix $\bm{U} \in \mathbb{R}^{10000\times10}$ is  given by the Gaussian distribution of $\mathcal{N}(0,1)$. The number of randomized space $s$ is set to be $s=n/10=1000$. 

\begin{figure}[!tbp]
\centerline{\includegraphics[width=9cm]{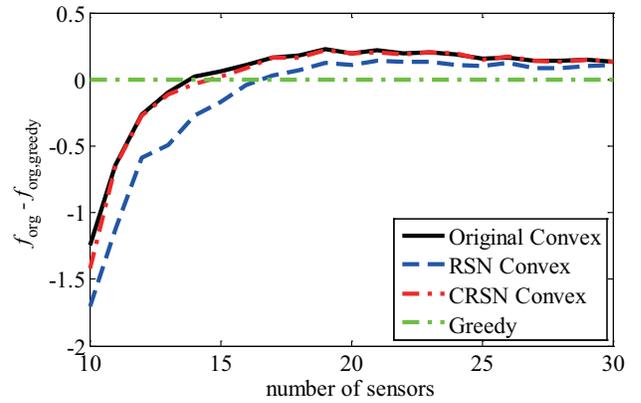}}
\caption{Difference of the results of $f_{\textrm{org}}$ by the convex methods from that by the greedy method.}
\label{fig:forgzoom} 
\end{figure}

First, the converged results of the sensor selection are discussed. Upper lines of Fig. \ref{fig:f} shows the results of $f$ calculated by the original, RSN, and CRSN methods. The results of $f$ are almost the same as each other in the present study. This shows that the RSN and CRSN methods work pretty well. Then, the results of $f_{\text{org}}$ which is based on the best-$p$ sensors of $\bm{w}$ are discussed, comparing with the result of the greedy method as a reference. Lower lines of Fig. \ref{fig:f} illustrates that the greedy method works better at $p\approx r$ as also discussed in Manohar et al.\cite{manohar2018data} and Saito et al.\cite{saito2019determinantbased}, while original convex method works better at $p > 15$ conditions. These are the characteristics of the original methods. This point is further clearly illustrated by taking the difference from the result of greedy method as shown in Fig. \ref{fig:forgzoom}. The results of RSN method are slightly worse than those of the original convex method at $p \approx r$, while those of CRSN are close to those of the original convex method. This discrepancy in $f_{\text{org}}$ is larger than that in $f$. This implicates that the slight difference in weight $\bm{w}$ leads to the difference of choice of sensors $\tilde{\bm{w}}$. This might be because the convergence is not perfectly obtained in the RSN and CRSN convex methods. However, results of RSN and CRSN convex methods are close to those of the original convex formulation at $p>15$ conditions, and they outperform the greedy method. 

\begin{figure}[!tbp]
\centerline{\includegraphics[width=9cm]{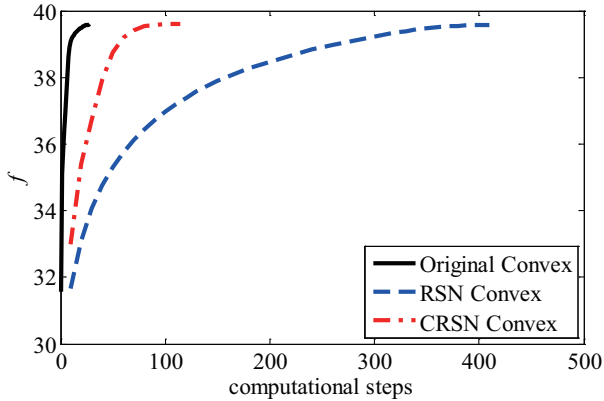}}
\caption{Step convergence of $f$ of the convex methods.}
\label{fig:step} 
\end{figure}

Then, Fig. \ref{fig:step} shows the variation of the objective function with computational steps at $p=20$. More computational steps are required for RSN and CRSN convex methods. The original convex method only requires 30 steps in median, while RSN and CRSN convex methods require 430 and 114 steps in median, respectively. If the convergence criterion is satisfied in succeeding $[n/s]=10$ steps, then the solution is judged to be converged and the effects of incorrect convergence criteria by the randomized algorithm are eliminated, as discussed before. This overhead might be reduced by further tuning up the algorithm. The RSN and CRSN convex methods require 20 and 10 times more computational step than the original convex method, respectively. These are expected feature of methods. 

\begin{figure}[!tbp]
\centerline{\includegraphics[width=9cm]{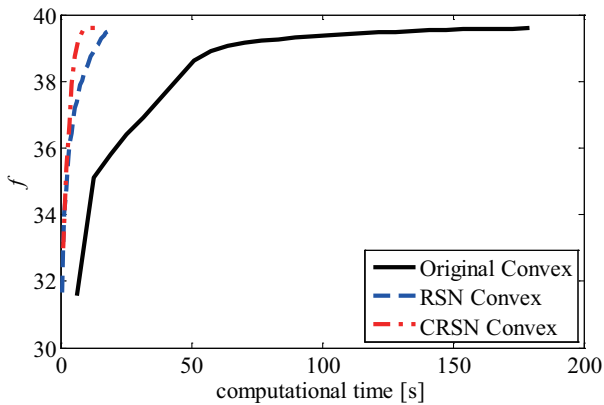}}
\caption{Time convergence of $f$ of the convex methods.}
\label{fig:time} 
\end{figure}

On the other hand, Fig. \ref{fig:time} illustrates the variation of the objective function with the computational time. Because one step of randomized methods is expected to be 1000 times faster, the convergence is obtained much faster than that by the original convex algorithm. Although the CRSN method requires a sorting procedure and has overhead as discussed before, convergence is accelerated by using the $\rho s$-largest weight sensors and it works slightly better than the RSN method. 

\begin{figure}[!tbp]
\centerline{\includegraphics[width=9cm]{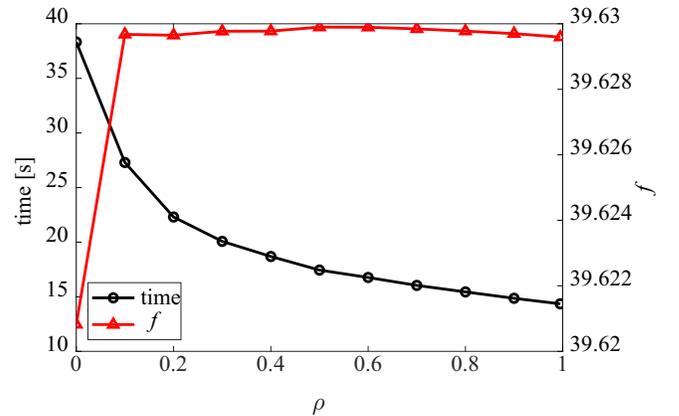}}
\caption{Effects of $\rho$ on the convergence value and the computational time in the CRSN convex method.}
\label{fig:CRSN} 
\end{figure}

\reviewerA{Finally, the effects of $\rho$ introduced for CRSN are investigated. The cases with $r=10$ and $p=20$ are calculated for 200 times with different random seeds, and the results are averaged. Fig. \ref{fig:CRSN} shows the converged value and the corresponding computational time against $\rho$. Here, $\rho=0$ is RSN and only this case does not have overhead of the 'sort' algorithm in computational time. Also, it should be noted that the range of converged values in this plot is very small. Interestingly, all the results of CRSN with different $\rho$ have better converged values and shorter computational times than those of RSN ($\rho=0$). Especially, the difference in converged values between CRSN is much smaller than the difference with RSN. This indicates that $\rho$ does not have strong effects on the converged value if CRNS is employed. In addition, the converged value of CRNS has the maximum at $\rho \approx 0.5$. This might be because elite sensors were sufficiently selected for update while the diversity is maintained in appropriate balance. Although the best $\rho$ value might be varied for the cases, we recommend $\rho=0.5$ for keeping the diversity of the sensors updated.}

\subsection{Data-Driven Sensor Selection Problem}
Here, we applied the present methods to the data-driven sensor selection problem. The sea temperature reconstruction problem is considered. See Refs. \cite{manohar2018data,saito2020data} for details of the problem. We have $\bm{U}\in \mathbb{R}^{44,000\times 10}$ sensor candidate of the observation of the temperature and the temperature distribution is reconstructed by estimating the strength of the proper orthogonal decomposition modes from the limited sparse observation. Here, $p$ is set to be 20 in this demonstration. Table \ref{tab:noaa} shows The results of the sensor selection time and resulting $f_{\text{org}}$. %Figure \ref{fig:noaa} shows the corresponding sensor locations. 
These results show that the computational time is much saved while keeping the performance of the sparse sensors.

% \begin{figure}[!tbp]
% \centerline{\includegraphics[width=10.0cm]{figs/NOAA.eps}}
%  \caption{Sensor location obtained by convex methods compared with the random selection.}
%  \label{fig:noaa} 
% \end{figure}

\begin{table}[!tbp]
\caption{Results of NOAA-SST problem. The same PC environment as in the previous study is employed. The random selection is average of 10,000 trials.}
\label{table}
\small
\setlength{\tabcolsep}{3pt}
\begin{tabular}{|c|c|c|c|}
\hline
Methods& 
$f_\textrm{org}$& 
computational time& computational steps\\
\hline
original convex    &-59.04 &26,002 s  &107 \\
RSN convex        &-59.28 &776 s   &685\\
CRSN convex    &-59.31 &384 s   &172\\
Random selection &   -82.79 & Not measured  &Not applicable\\
\hline
\end{tabular}
\label{tab:noaa}
\end{table}

\section{Conclusions}
\label{sec:con}
In the present paper, the randomized subspace Newton method is successfully applied to the convex relaxation formulation of the sensor selection. The resulting method is effective for the problem with many sensor candidates as is often the case of data-driven sensor selection problems. Although the convex formulation with the randomized algorithm is still slower than greedy method, the better performance and the further extension to the complex objective function including constraints is expected by using the framework proposed in the present study. 

\section*{Acknowledgment}
This work was supported by JST CREST (JPMJCR1763) and ACT-X (JPMJAX20AD), Japan.
\clearpage

\bibliographystyle{IEEEtran}
\bibliography{xaerolab}

\end{document}